\documentclass[aps,letter,prd,fleqn,showpacs,showkeys,twocolumn,nofootinbib,
preprintnumbers,superscriptaddress]{revtex4-1}
\usepackage{amsmath,amsfonts,amssymb,amscd,amsxtra,amsthm}
\usepackage{graphicx}  
\usepackage{epstopdf}
\usepackage{color}
\usepackage{dcolumn}  
\usepackage{bm}          
\usepackage{multirow}
\usepackage{slashed}
\usepackage[normalem]{ulem}
\usepackage[utf8]{inputenc}

\begin{document}

\title{Landscape of heavy baryons \\ from the perspective of the chiral quark-soliton model}

\author{\framebox[1.05\width]{Maxim V. Polyakov}}
\email{On August 25, 2021 Maxim Polyakov passed away prematurely while
 working on the above paper. This article is based on the collaboration
with Maxim and is dedicated to the memory of an outstanding physicist and a very good friend.}

\affiliation{Institut f\"ur Theoretische Physik II, Ruhr-Universit\"at
  Bochum, D--44780, Bochum, Germany}

\author{Micha{\l{}} Prasza{\l{}}owicz}\email{michal.praszalowicz@uj.edu.pl}
\affiliation{Institute of Theoretical Physics, Jagiellonian
  University, {\L}ojasiewicza 11, 30-348 Krak{\'o}w, Poland}

\begin{abstract} 
We employ the chiral quark-soliton model and the heavy quark symmetry to
describe spectra of charm and beauty baryons. 
Heavy baryons can be classified according to the SU(3) representations of the light sector.
We argue 
 that recently discovered $\Xi_b$ states can be interpreted as negative-parity excited
antitriplets or sextets, and the $\Sigma_b$ states as negative-parity sextets.
Consequences of such assignments 
for the decay patterns are discussed and also predictions of masses of the yet unmeasured 
sextet members are given.
\end{abstract}



\maketitle

\section{Introduction}

Recent discoveries of heavy baryons, \emph{i.e.} baryons with one heavy quark
$Q=c$ or $b$\footnote{Doubly heavy baryons, like $\Xi_{cc}$
\cite{SELEX:2004lln,LHCb:2017iph}, are beyond scope of this paper.} renewed
interest in heavy baryon spectroscopy. In 2017 the LHCb Collaboration
announced five new excited $\Omega_{c}^{0}$ states, two of them of a very
small width~\cite{LHCb:2017uwr}, which were confirmed by the Belle
Collaboration~\cite{Belle:2017ext} in 2018. Further analysis of the decay
modes and possible spin assignment of these states has been published 
recently (Ref.~\cite{LHCb:2021ptx}). Furthermore in 2018 LHCb announced new
$b$ baryons: the isospin doublet of $\Xi_{b}(6227)$~\cite{LHCb:2018vuc},
charged members of $\Sigma_{b}(6097)$ multiplet~\cite{LHCb:2018haf} and two
nearly degenerate $\Lambda_{b}$ baryons~\cite{LHCb:2019soc} at 6146~MeV/$c^{2}$ and 6152
MeV/$c^{2}$, which are today interpreted as $1/2^{+}$ and $3/2^{+}$ spin
states~\cite{PDG}. However, the interpretation of these states as excited
$\Sigma_{b}^{0}$ states cannot be excluded. In 2020 LHCb reported four
$\Omega_{b}^{-}$ excited states~\cite{LHCb:2020tqd}, which are presently
considered to be only one star resonances~\cite{PDG}. If confirmed, they would
pose, as we will discuss later, a challenge for theoretical interpretation.
Next, LHCb reported a new $\Lambda_{b}^{0}(6072)$ state~\cite{LHCb:2020lzx}, a
new $\Xi_{b}^{0}(6227)$ state~\cite{LHCb:2020xpu} and two other $\Xi_{b}^{0}$
baryons~\cite{LHCb-talk} at 6327~MeV/$c^{2}$ and 6333~MeV/$c^{2}$. Finally, the CMS
Collaboration reported a $\Xi_{b}^{-}(6100)$ baryon interpreted as, presumably,
a $3/2^{-}$ spin state \cite{CMS:2021rvl}. All these new baryons, as well as the ones found earlier
and included in the PDG~\cite{PDG}, are listed in Sec.~\ref{sec:exp} for the
convenience of the reader.

There is a wealth of literature devoted to the theoretical description of heavy baryons,
for a complete list of references for charmed baryons we refer the reader to a recent
review by Hai-Yang Cheng \cite{Cheng:2021qpd}. Likewise an extensive bibliography
for bottom baryons can be found in Ref.~\cite{Oudichhya:2021yln}. The approach that
we will advocate in the present paper is closely related to the approaches based on heavy quark
effective theory~\cite{Yamaguchi:2014era},
quark-diquark models
 \cite{Ebert:2007nw,Ebert:2011kk}
and  the constituent quark model \cite{Bijker:2020tns,gaps_Chen,Lv:2020qi,Chen:2016iyi,Kakadiya:2021jtv,Yoshida:2015tia,Cheng:2015naa},

In the present paper we will continue our earlier analysis 
\cite{Yang:2016qdz,Praszalowicz:2018uoa,Kim:2017jpx,Kim:2017khv}
based on the Chiral-Quark Soliton Model
($\chi$QSM) \cite{Diakonov:1987ty,Christov:1995vm,Alkofer:1994ph,Petrov:2016vvl}
where baryons are viewed as rotational excitations of the Chiral-Quark Soliton Model, which
is a good description of the light baryonic system in the large $N_c$ limit \cite{Witten:1979kh}.
Heavy
baryons are constructed by removing one  quark from the $N_c$ light quarks and replacing
it by a heavy quark $Q$. In the large $N_c$ limit the light sector is hardly changed by such a replacement,
and one can describe both ground state and excited positive parity baryons rather successfully 
\cite{Yang:2016qdz,Praszalowicz:2018uoa,Kim:2017jpx,Kim:2017khv}. Excitations appear due to the chiral
rotation of the soliton and are therefore analogous to the diquark excitations (so-called $\rho$ modes) in the
quark language.

At this point it is useful to compare the present model with very popular approach
to heavy baryons based on a diquark-heavy~quark dynamics (see {\em e.g.}, \cite{Cheng:2015naa}). In such models
one distinguishes two types of excitations: inner diquark excitations, referred often to
as $\rho$ modes, and diquark-heavy~quark excitations referred to as  $\lambda$ modes. From this
point of view the soliton model described above corresponds only to the rotational  $\rho$ modes. This can be justified by
invoking large $N_c$ arguments, see Sec.~\ref{sec:chiQSM}, on which the whole approach used here is based.
Therefore in the following we in fact test a hypothesis, whether heavy baryon masses can be explained
if $\lambda$ modes are neglected. Further support for neglecting the $\lambda$ modes comes from the bound-state 
approach to heavy baryons in the Skyrme model and is discussed in Sec.~\ref{ssec:Skyrme}.
Phenomenologically, however, $\lambda$ modes may of course play a non-negligible role.

Even with $\lambda$ modes neglected the heavy baryon spectrum is very rich. If, furthermore, $\lambda$ modes
and also radial excitations are included, one predicts a densely populated spectrum in the mass range already 
scanned by different experiments. Since the experimental evidence is much less abundant, the
large $N_c$ counting discussed above may serve as an effective Occam's razor.

The price  for such a truncation is, in principle, rather small accuracy of numerical predictions, our approach is certainly
more qualitative than quantitative. Nevertheless some predictions are very accurate, for example the sum rule
involving $\Omega_c$ states -- see caption in Table~\ref{tab:6_b} and Eq.~(22) in Ref.~\cite{Yang:2016qdz}.

We show that recently discovered $\Xi_b$ states can be interpreted as negative-parity excited
antitriplets or sextets, and the $\Sigma_b$ states as sextets. We discuss the consequences of 
this assignment for the decay patterns and for the charm sector.

The paper is organized as follows: In Sec.~\ref{sec:exp} we review the present
experimental situation in heavy baryon sector and summarize phenomenological
expectations based on the naive extrapolation of the $c$ sector to the $b$
sector. Next, in Sec.~\ref{sec:chiQSM}, we briefly introduce the Chiral
Quark-Soliton Model ($\chi$QSM) with emphasis on its extension to heavy
baryons. In Sec.~\ref{sect:pheno} we analyze phenomenological consequences of the
$\chi$QSM for the ground states including exotica
(Sec.~\ref{ssec:pheno_ground}), excited negative-parity antitriplets (Sec.~\ref{ssec:pheno_excited3bar})
and sextets (Sec.~\ref{subs:6prim}). In the latter case we discuss possible SU(3) assignments 
of $\Sigma_{b}(6097)$, $\Xi_{b}(6327)$, and $\Xi_{b}(6333)$  to fix model parameters and  we
discuss consequences of these assignments for the charm sector.  Next in Sec.~\ref{sect:disc} we try to understand
decay patterns and analyze how they impact some of the recent experiments.
Finally we conclude in Sec.~\ref{sec:conclusions}.

\section{Experiment }
\label{sec:exp}

Ground state heavy baryons can be conveniently classified according to the
SU(3) structure of the light subsystems (diquarks), which can be in the spin 0
antitriplet and spin-1 sextet. The same group structure follows from the
$\chi$QSM \cite{Yang:2016qdz}. Adding one heavy quark to the light subsystem one ends up 
with spin-$1/2^{+}$ antitriplet and two sextets of spin $1/2^{+}$ and $3/2^{+}$, which
are hyperfine split and the splittings are proportional to the inverse of the
heavy quark mass. These structures are fully confirmed experimentally, as can
be seen from Tables~\ref{tab:3bar_c} and \ref{tab:3bar_b}. In these Tables we
also display the first orbitally excited negative-parity antitriplets that form
two hyperfine split multiplets of spin $1/2^{-}$ and $3/2^{-}$. All members of
these two antitriplets in the charm sector have been known for a long time.
 Until recently only $\Lambda_{b}^{0}$'s have been measured in the $b$ sector.
In 2021 one $\Xi_{b}^{-}$ state, presumably $3/2^{-}$, was observed by the CMS
Collaboration at CERN \cite{CMS:2021rvl}.

Looking at masses, both in the ground state and the excited antitriplets, we
observe with accuracy~$< 15 \%$ equal splittings between $\Xi_{Q}$ and
$\Lambda_{Q}$ states. Indeed (in MeV/$c^{2}$),
\begin{align}
\delta_{c}^{1/2^{+}}(\Xi-\Lambda) & \approx182,~~\delta_{c}^{1/2^{-}}(\Xi
-\Lambda)\approx201,\notag \\ 
\delta_{c}^{3/2^{-}}(\Xi-\Lambda) & \approx190,~~\delta
_{b}^{1/2^{+}}(\Xi-\Lambda)\approx175 \, .\label{eq:delta3barpheno}%
\end{align}
Equal mass splittings in each parity multiplet, independently of the heavy
quarks involved, follow from the SU(3) and heavy quark symmetries. However the
equality of splittings in different parity multiplets is not obvious. As we
have shown in Ref.~\cite{Kim:2017jpx} this is a prediction of the $\chi$QSM.
Since we expect heavy quark symmetry to work better in the $b$ sector, we
immediately get predictions for the isospin averaged masses of excited $\Xi_{b}$'s,
\begin{align}
\Xi_{b}^{1/2^{-}}=\Lambda_{b}^{1/2^{-}}+\delta_{b}^{1/2^{+}}(\Xi-\Lambda) &
\approx6087~\text{MeV}/c^{2}\, ,\nonumber\\
\Xi_{b}^{3/2^{-}}=\Lambda_{b}^{3/2^{-}}+\delta_{b}^{1/2^{+}}(\Xi-\Lambda) &
\approx6095~\text{MeV}/c^{2}\, .\label{eq:Xibpredictions}%
\end{align}
First equation is a prediction, and the second one
is in perfect
agreement with recent CMS~\cite{CMS:2021rvl} finding\footnote{Note that CMS
has measured only $\Xi_{b}^{-}$, so $\Xi_{b}^{0}$, which we expect to be lower
in mass, is still to be found. Therefore, the average isospin mass is expected
to be lower than 6100~MeV/$c^2$.}, see Table \ref{tab:3bar_b}.

Similar pattern is observed for the ground-state sextets, both in charm (in MeV/$c^{2}$)
\begin{align}
\delta_{c}^{1/2^{+}}(\Xi-\Sigma) & \approx124,~~\delta_{c}^{1/2^{+}}(\Omega
-\Xi)\approx117,\notag \\
 \delta_{c}^{3/2^{+}}(\Xi-\Sigma) & \approx128,~~\delta
_{c}^{1/2^{+}}(\Omega-\Xi)\approx120\label{eq:6cpheno}%
\end{align}
and in beauty sectors (in MeV/$c^{2}$),
\begin{align}
\delta_{b}^{1/2^{+}}(\Xi-\Sigma) & \approx122,~~\delta_{b}^{1/2^{+}}(\Omega
-\Xi)\approx111,\notag \\
\delta_{b}^{3/2^{+}}(\Xi-\Sigma) & \approx121 \,
.\label{eq:6bpheno}%
\end{align}
Again, we see the independence of the splittings from the mass of the heavy
quark. Unfortunately, as we shall see in the following, the $\chi$QSM predicts
that splittings in the excited sextets should be different~\cite{Kim:2017jpx}.

\begin{table}[h]
\centering
\begin{tabular}
[c]{|c|c|cc|}\hline
$S^{P}$ & $\Lambda_{c}^{+}$ & \multicolumn{2}{c|}{$\Xi_{c}^{+,0} $}\\\hline
$1/2^{+}$ & $2286.46\pm0.14 $ & $%
\begin{array}
[c]{c}%
2467.71\pm023\\
2470.44 \pm0.28
\end{array}
$ & (2469)\\\hline
$1/2^{-} $ & $2592.25\pm0.28 $ & $%
\begin{array}
[c]{c}%
2791.90 \pm0.50\\
2793.90 \pm0.50
\end{array}
$ & (2793)\\\hline
$3/2^{-} $ & $2628.11 \pm0.19 $ & $%
\begin{array}
[c]{c}%
2816.51 \pm0.25\\
2819.79 \pm0.60
\end{array}
$ & (2818)\\\hline
\end{tabular}
\caption{Ground-state and excited charm baryons in the SU(3) antitriplet. All
listed baryons are PDG 3-star resonances~\cite{PDG}. Masses are in MeV/$c^{2}%
$. Entries in parenthesis denote isospin averages used for numerical
calculations.}%
\label{tab:3bar_c}%
\end{table}

\begin{table}[h]
\centering
\begin{tabular}
[c]{|c|c|cc|}\hline
$S^{P}$ & $\Lambda_{b}^{0}$ & \multicolumn{2}{c|}{$\Xi_{b}^{-,0} $}\\\hline
$1/2^{+}$ & $5619.60\pm0.17 $ & $%
\begin{array}
[c]{c}%
5797.0\pm0.6\\
5791.9\pm0.5
\end{array}
$ & (5795)\\\hline
$1/2^{-} $ & $5912.19\pm0.17 $ & $%
\begin{array}
[c]{c}%
... \\
...
\end{array}
$ & \\\hline
$3/2^{-} $ & $5920.08\pm0.17 $ & $%
\begin{array}
[c]{c}%
6100.3^{*}\pm0.64\\
...
\end{array}
$ & (6100)\\\hline
\end{tabular}
\caption{Ground-state and excited beauty baryons in the SU(3) antitriplet.
All listed baryons are PDG 3-star resonances~\cite{PDG}, except for $\Xi
_{b}^{-}$ marked with a star, which has been reported last year by
CMS~~\cite{CMS:2021rvl} and is not included in 2021 PDG. Masses are in
MeV/$c^{2}$. Entries in parenthesis denote isospin averages used for numerical
calculations. Isospin partners not yet measured are marked by dots.}%
\label{tab:3bar_b}%
\end{table}

\begin{widetext}

\begin{table}[h]
\centering
\begin{tabular}
[c]{|c|cc|cc|c|}\hline
$S^{P} $ & \multicolumn{2}{c|}{$\Sigma_{c}^{++,+,0,}$} &
\multicolumn{2}{c|}{$\Xi^{\prime\, +,0}_{c}$} & $\Omega^{0}_{c}$\\\hline
$1/2^{+}$ & $%
\begin{array}
[c]{c}%
2453.97\pm0.14\\
2452.90 \pm0.40\\
2453.75 \pm0.14
\end{array}
$ & (2454) & $%
\begin{array}
[c]{c}%
2578.20 \pm0.50\\
2578.70 \pm0.50
\end{array}
$ & (2578) & $2695.20 \pm1.70$\\\hline
$3/2^{+} $ & $%
\begin{array}
[c]{c}%
2518.41 \pm0.20\\
2517.50 \pm2.30\\
2518.48 \pm0.20
\end{array}
$ & (2518) & $%
\begin{array}
[c]{c}%
2645.10 \pm0.30\\
2646.16 \pm0.25
\end{array}
$ & (2646) & $2765.90 \pm2.00 $\\\hline
\end{tabular}
\caption{Ground-state charm baryons in the SU(3) sextet. All listed baryons
are PDG 3-star resonances~\cite{PDG}. Masses are in MeV/$c^{2}$. Entries in
parenthesis denote isospin averages used for numerical calculations. }%
\label{tab:6_c}%
\end{table}

\begin{table}[h]
\centering
\begin{tabular}
[c]{|c|cc|cc|c|}\hline
$S^{P} $ & \multicolumn{2}{c|}{$\Sigma_{b}^{+,0,-}$} &
\multicolumn{2}{c|}{$\Xi^{\prime\, -,0}_{b}$} & $\Omega^{-}_{b}$\\\hline
$1/2^{+}$ & $%
\begin{array}
[c]{c}%
5810.56\pm0.25\\
...\\
5815.64\pm0.27
\end{array}
$ & (5813) & $%
\begin{array}
[c]{c}%
5935.02\pm0.05\\
...
\end{array}
$ & (5935) & $6046.1\pm1.7$\\\hline
$3/2^{+} $ & $%
\begin{array}
[c]{c}%
5830.32\pm0.27\\
...\\
5834.74\pm0.30
\end{array}
$ & (5833) & $%
\begin{array}
[c]{c}%
5952.30\pm0.60\\
5955.33\pm0.12
\end{array}
$ & (5954) & $6076.8\pm2.2^{*} $\\\hline
\end{tabular}
\caption{Ground-state beauty baryons in the SU(3) sextet. All listed baryons
are PDG 3-star resonances~\cite{PDG}, except for $\Omega_{b}(3/2^{+})$ marked
with a star, which has not been measured; the mass is a prediction from the sum rule
derived in Ref.~\cite{Yang:2016qdz}. Masses are in MeV/$c^{2}$. Entries in
parenthesis denote isospin averages used for numerical calculations. Isospin
partners not yet measured are marked by dots. }%
\label{tab:6_b}%
\end{table}
\end{widetext}

\begin{table}[h]
\centering
\begin{tabular}
[c]{|cc|ccc|}\hline
$S^{P}$ & $\Lambda_{c}^{+}$ & $S^{P}$ & \multicolumn{2}{c|}{$\Xi^{+,0}_{c}$%
}\\\hline
? & $2766.60 \pm2.40 $ & \multirow{2}{*}{?} & $...$ &
\multirow{2}{*}{$(2923)$}\\
$3/2^{+} $ & $\mathbf{2856.10 \pm5.60}$ &  & $2923.04\pm0.25$ & \\\cline{3-5}%
$5/2^{+} $ & $\mathbf{2881.63\pm0.24}$ & \multirow{2}{*}{?} & $2942.30\pm4.40$
& \multirow{2}{*}{$(2940)$}\\
$3/2^{-}$ & $\mathbf{2939.60\pm1.50} $ &  & $2938.55\pm0.22$ & \\\cline{3-5}
&  & \multirow{2}{*}{?} & $\mathbf{2964.30 \pm1.50}$ &
\multirow{2}{*}{$(2966)$}\\
&  &  & $\mathbf{2967.10 \pm1.70}$ & \\\cline{3-5}
&  & \multirow{2}{*}{?} & $\mathbf{3055.90\pm0.40} $ &
\multirow{2}{*}{$(3056)$}\\
&  &  & $...$ & \\\cline{3-5}
&  & \multirow{2}{*}{?} & $\mathbf{3077.20\pm0.40}$ &
\multirow{2}{*}{$(3079)$}\\
&  &  & $\mathbf{3079.90\pm1.40}$ & \\\cline{3-5}
&  & \multirow{2}{*}{?} & $3122.90\pm1.30$ & \multirow{2}{*}{$(3123)$}\\
&  &  & $...$ & \\\hline\hline
$S^{P}$ & $\Omega_{c}^{0}$ & $S^{P}$ & \multicolumn{2}{c|}{$\Sigma^{
++,+,0}_{c}$}\\\hline
? & $\mathbf{3000.41\pm0.22}$ & \multirow{3}{*}{?} & $\mathbf{2801.\pm6.0}$ &
\multirow{3}{*}{$(2800)$}\\
? & $\mathbf{3050.20 \pm0.13}$ &  & $\mathbf{2792.\pm14.}$ & \\
? & $\mathbf{3065.46 \pm0.28}$ &  & $\mathbf{2806. \pm7.0} $ & \\\cline{3-5}%
? & $\mathbf{3090.00 \pm0.50}$ &  &  & \\
? & $\mathbf{3119.10 \pm0.90}$ &  &  & \\
? & $3188.00\pm13.0$ &  &  & \\\hline
\end{tabular}
\caption{Charm baryons with unknown SU(3) assignment. Entries in bold face
denote 3-star PDG resonances~\cite{PDG}, other entries are 1-star or 2-star
resonances omitted from the summary listings, except for $\Omega^{0}%
_{c}(3188)$ taken from~\cite{LHCb:2021ptx}. Masses are in MeV/$c^{2}$. Isospin
partners not yet measured are marked by dots. Entries in parenthesis denote
isospin averages used for numerical calculations. Experimental errors are for
orientation only (typically the largest PDG error is listed).}%
\label{tab:unknown_c}%
\end{table}

\newpage

\begin{table}[h]
\centering
\begin{tabular}
[c]{|cc|ccc|}\hline
$S^{P}$ & $\Lambda_{b}^{0}$ & $S^{P}$ & \multicolumn{2}{c|}{$\Xi^{-,0}_{b}$%
}\\\hline
? & $6072.3\pm2.9 $ & \multirow{2}{*}{?} & $\mathbf{6227.9\pm0.9} $ &
\multirow{2}{*}{$(6227)$}\\
$3/2^{+} $ & $\mathbf{6146.2\pm0.4}$ &  & $\mathbf{6226.8\pm1.5 }$ &
\\\cline{3-5}%
$5/2^{+}$ & $\mathbf{6152.5\pm0.4}$ & \multirow{2}{*}{?} & $... $ &
\multirow{2}{*}{$(6327)$}\\
&  &  & ${6327.28^{*} \pm0.33}$ & \\\cline{3-5}%
$$ &  & \multirow{2}{*}{?} & $... $ & \multirow{2}{*}{$(6333)$}\\
&  &  & ${ 6332.69^{*} \pm0.31}$ & \\\hline\hline
$S^{P}$ & $\Omega_{b}^{-}$ & $S^{P}$ & \multicolumn{2}{c|}{$\Sigma^{
+,0,-}_{b}$}\\\hline
? & $6315.6 \pm0.58$ & \multirow{3}{*}{?} & $\mathbf{6095.8 \pm1.7}$ &
\multirow{3}{*}{$(6097)$}\\
? & $6330.3 \pm0.58$ &  & $...$ & \\
? & $6339.7 \pm0.58$ &  & $\mathbf{6098.0 \pm1.7 }$ & \\\cline{3-5}%
? & $6349.8 \pm0.64$ &  &  & \\\hline
\end{tabular}
\caption{Beauty baryons with unknown SU(3) assignment. Entries in bold face
denote 3-star PDG resonances~\cite{PDG}, other entries are 1-star or 2-star
resonances omitted from the summary listings, except for two $\Xi_{b}^{0}$
resonances marked with a star, which have been reported this year by
LHCb~\cite{LHCb-talk} and are not included in 2021 PDG. Masses are in
MeV/$c^{2}$. Isospin partners not yet measured are marked by dots. Entries
in parenthesis denote isospin averages used for numerical calculations.
Experimental errors are for orientation only (typically the largest PDG error
is listed).}%
\label{tab:unknown_b}%
\end{table}

It is interesting to look at the splittings between different heavy quark
multiplets. Here the simplest comparison can be made for the ground state
triplets
\begin{equation}
\delta(\Lambda_{b}-\Lambda_{c})= 3333,~~\delta(\Xi_{b}-\Xi_{c})=3326 \,
,\label{eq:bcdiff}%
\end{equation}
which corresponds to the difference $m_{b}-m_{c}$ in MeV/$c^{2}$. Notably this
is $400 \div200$~MeV/$c^{2}$ higher than the PDG~\cite{PDG} value either for
the $\overline{\text{MS}}$ or the pole mass, respectively. Similar mass
differences for excited antitriplets and for sextets require taking spin
effects into account, and we relegate this to the next sections. Spin
splittings in the charm sector are of the order of $\pm17$~MeV/$c^{2}$ and in
the beauty case $\pm4$ MeV/$c^{2}$. Neglecting spin effects we expect
the $b$-baryon spectrum to be a copy of  $c$-baryons shifted by approximately
$3330\pm20$ MeV/$c^{2}$. Given recent measurements of five $\Omega_{c}$'s that
fall between 3000 MeV/$c^{2}$ and 3120 MeV/$c^{2}$, one would expect similar structure in
the $b$ sector in an interval between $(6330 - 6450) \pm20$ MeV/$c^{2}$. One
sees indeed four $\Omega_{b}$ candidates (see Table~\ref{tab:unknown_b}) but
in a much narrower interval of 35 MeV/$c^{2}$ only. One should, however,
remember that $\Omega_{b}$ states are one-star resonances and are not listed
in the PDG summary listings~\cite{PDG}. For possible interpretation of these states
as doubly-strange $bss$ states see Ref.~\cite{Karliner:2020fqe} and also \cite{Wang:2020pri}.

\section{Chiral Quark Soliton Model}

\label{sec:chiQSM}

In this section we recapitulate shortly the chiral quark-soliton model, for
more detailed discussion we refer the reader to the original works
\cite{Diakonov:1987ty} and to the reviews of
Refs.~\cite{Christov:1995vm,Alkofer:1994ph,Petrov:2016vvl} (and references
therein). The $\chi$QSM is based  on an old argument by
Witten~\cite{Witten:1979kh}, that in the limit of a large number  of colors
($N_{c} \rightarrow\infty$),  $N_{c}$ relativistic valence quarks generate
chiral mean fields represented by a distortion of  the Dirac sea. Such
distortion interacts with the valence quarks changing their wave function,
which in turn modifies the sea until stable configuration is reached. This
configuration called  \emph{chiral soliton} corresponds to the solution of the
Dirac equation for the constituent quarks (with gluons integrated out) in the
mean-field approximation where the mean fields respect so called
{hedgehog} symmetry. {Hedgehog} symmetry follows from the fact that
it is impossible to construct a pseudoscalar field that changes sign under
inversion of coordinates, which would be compatible with the
SU(3)$_{\mathrm{flav}}\times$SO(3) space symmetry. It has been shown that the
{hedgehog} symmetry, which is smaller than SU(3)$_{\mathrm{flav}}\times
$SO(3), leads to the correct baryon spectrum (see below).

\begin{figure}[h]
\centering
\includegraphics[width=9.0cm]{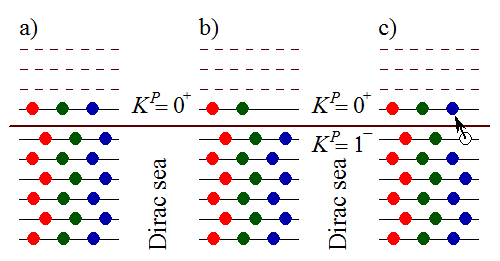} \vspace{-0.2cm}
\caption{Schematic pattern of light quark levels in a self-consistent soliton
configuration. In the left panel all sea levels are filled and $N_{c}$ ($=3$ in
the Figure) valence quarks occupy the $K^{P}=0^{+}$ lowest positive energy
level. Unoccupied positive energy levels are depicted by dashed lines. In the
middle panel one valence quark has been stripped off, and the soliton has to
be supplemented by a heavy quark not shown in the Figure. In the right panel a
possible excitation of a sea level quark, conjectured to be $K^{P}=1^{-}$, to
the valence level is shown, and again the soliton has to couple to a heavy
quark. (Figure from Ref.~\cite{Kim:2017jpx}.)}%
\label{fig:levels}%
\end{figure}

In vacuum quarks are characterized by two independent SU(2) symmetries: spin
$\bm{S}$ and isospin $\bm{T}$. In the soliton configuration, due to the
{hedgehog} symmetry, neither spin, nor isospin are {good} quantum
numbers. Instead a {grand spin} $\bm{K}=\bm{S}+\bm{T}$ is a {good}
quantum number. Solutions of the Dirac equations are therefore labeled by
$K^{P}$ quantum numbers ($P$ standing for parity). The ground state
configuration corresponds to the fully occupied $K^{P}=0^{+}$ valence level,
as shown in Fig.~\ref{fig:levels}.a. Therefore the soliton does not carry
definite quantum numbers except for the baryon number resulting from the
valence quarks.

Spin and isospin appear when the rotations in space and flavor are quantized.
This procedure results in a {collective} Hamiltonian analogous to the one
of a quantum mechanical symmetric top. There are two conditions that the {collective} wave
functions have to satisfy:

\begin{itemize}
\item allowed SU(3) representations must contain states with hypercharge
$Y^{\prime}=N_{c}/3$,

\item the isospin $\bm{T}^{\prime}$ of the states with $Y^{\prime}=N_{c}/3$
couples with the soliton spin $\bm{J}$ to a singlet: $\bm{T}^{\prime
}+\bm{J}=0$.
\end{itemize}

Such a configuration has been used to describe octet and decouplet of light
baryons, and is not of interest for us here.

Instead, we will focus on the configuration depicted in in
Fig.~\ref{fig:levels}.b, when one light quark has been removed from the
valence level. Since the valence level has grand spin $K=0$, such
configuration does not carry any quantum numbers other than the baryon number
that is $2/3$, or strictly speaking $(N_{c}-1)/N_{c}$. Such a soliton has to be
supplemented by a heavy quark to form a baryon with baryon number equal one.
Following \cite{Diakonov:2013qta}, it has been shown in
Ref.~\cite{Yang:2016qdz} that such a model satisfactorily fits experimental data.

In the large $N_{c}$ limit light sector both in light and heavy
baryons is described essentially by the same mean field. The only difference
is now in the quantization condition:

\begin{itemize}
\item allowed SU(3) representations must contain states with hypercharge
$Y^{\prime}=(N_{c}-1)/3$,

\item the isospin $\bm{T}^{\prime}$ of the states with $Y^{\prime}$ couples
with the soliton spin $\bm{J}$ to a singlet: $\bm{T}^{\prime}+\bm{J}=0$.
\end{itemize}

\begin{figure}[h]
\centering
\includegraphics[height=4cm]{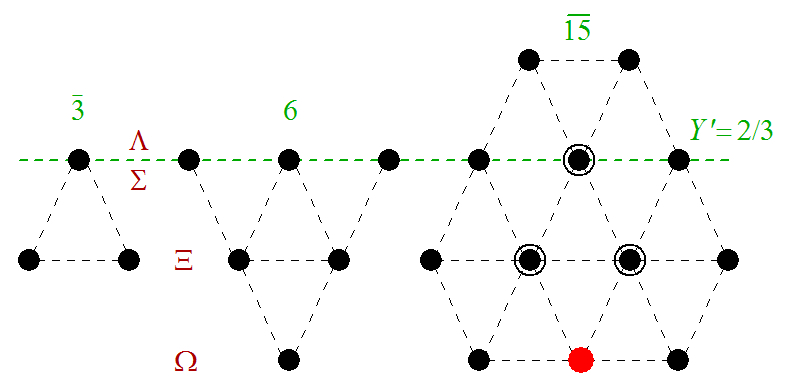}\caption{Rotational band of a
soliton with one valence quark stripped off. Soliton spin corresponds to the
isospin $T^{\prime}$ of states on the quantization line $Y^{\prime}=2/3$. We
show three lowest allowed representations: antitriplet of spin 0, sextet of
spin 1 and the lowest exotic representation $\overline{\mathbf{15}}$ of spin 1
or 0. Heavy quarks have to be added. Red dot corresponds to the exotic
$\Omega_{Q}(T_{3}=0)$ discussed in \cite{Kim:2017jpx}. (Figure from
Ref.~\cite{Praszalowicz:2018uoa}.)}%
\label{fig:irreps}%
\end{figure}

The lowest allowed SU(3) representations for such configurations are
$\overline{\mathbf{3}}$ of spin 0 and  ${\mathbf{6}}$ of spin 1, as in the
quark model. They are shown in Fig.~\ref{fig:irreps}
together with exotic $\overline{\mathbf{15}}$ that corresponds to
heavy pentaquarks \cite{Kim:2017jpx}.
Therefore, heavy baryons,
that are constructed from the soliton and a heavy quark form an SU(3)
antitriplet of spin 1/2 and two sextets of spin 1/2 and 3/2 that are subject
to a hyperfine splitting.

Finally, following Ref.~\cite{Kim:2017jpx} we shall conjecture that the first
occupied sea level in Fig.~\ref{fig:levels} is $K=1^{-}$. If we excite one
quark from this shell to the free valence level, the soliton will have not
only the baryon number but also $K$ corresponding to the hole in the sea. In
this case, shown in Fig.~\ref{fig:levels}.c, baryon parity is negative and the
quantization condition reads:

\begin{itemize}
\item allowed SU(3) representations must contain states with hypercharge
$Y^{\prime}=(N_{c}-1)/3$,

\item the isospin $\bm{T}^{\prime}$ of the states with $Y^{\prime}%
=(N_{c}-1)/3$ couples with the soliton spin $\bm{J}$ as follows:
$\bm{T}^{\prime}+\bm{J}=\bm{K}$, where $\bm{K}$ is the grand spin of the
unoccupied sea level.
\end{itemize}

Therefore the rotational bands are the same as in Fig.~\ref{fig:irreps} with,
however, different spin assignments.

For $\overline{\mathbf{3}}$ $T^{\prime}=0$, however $K=1$, so the soliton has spin $J=1$
and we expect two hyperfine split heavy baryon antitriplets of spin-1/2 and
spin-3/2. As discussed in Sec.~\ref{sec:exp} this is indeed experimentally the case.

\begin{figure}[h]
\centering
\includegraphics[scale=0.550]{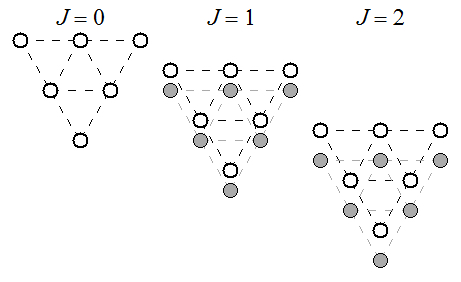} \caption{Excited heavy baryons
belonging to the SU(3) sextets.}%
\label{fig:e6}%
\end{figure}

For $\mathbf{6}$ $T^{\prime}=1$, which for $K=1$ gives the soliton spin
$J=0,1,2$ in direct analogy to the total angular momentum of the light
subsystem in the quark model. By adding one heavy quark we end up with five
possible excitations, which have the following total spin $S$: for $J=0$: $S=1/2$,
for $J=1$: $S=1/2$ and 3/2, and for $J=2$: $S=3/2$ and 5/2. States
corresponding to the same $J$ are subject to small hyperfine splittings as
depicted in Fig.~\ref{fig:e6}. All these states have negative-parity.

The formula for the soliton mass in the chiral limit for the states in the
SU(3) representation $\mathcal{R}$ has been derived in
Ref.~\cite{Diakonov:2013qta} and reads:

\begin{widetext}
\begin{align}
\mathcal{M}^{(K)} =M^{(K)}_{\text{sol}} & +\frac{1}{2I_{2}}\left[
C_{2}(\mathcal{R})-T^{\prime}(T^{\prime}+1)-\frac{3}{4}Y^{\prime2}\right] \label{rotmass} \\
& +\frac{1}{2I_{1}}\left[  (1-a_{K})T^{\prime}(T^{\prime}+1)\frac{{}}{{}}
+a_{K}J(J+1)-a_{K}(1-a_{K})K(K+1)\;\right] \nonumber%
\end{align}
\end{widetext} 
where $C_{2}(\mathcal{R})$ stands for the SU(3) Casimir operator.
$M_{\text{sol}}^{(K)}\sim N_{c}$ denotes classical soliton mass, $I_{1,2}$
represent moments of inertia and $a_{K}$ is a parameter that takes into
account one-quark excitation. Although all these parameters can, in
principle, be calculated in a specific model, we shall follow here a so-called
{model-independent} approach introduced in the context of the Skyrme
model in Ref.~\cite{Adkins:1984cf}, where all the parameters are extracted from
the experimental data. This approach has been used in the context of five
excited $\Omega_{c}$ states in Ref.~\cite{Kim:2017jpx}.

At this point it is important to note that Hamiltonian (\ref{rotmass}) corresponds to the excitations of the light sector only,
{\em i.e.} to the so called $\rho$ modes in the quark model language. Such excitations are dominant in the large $N_c$ limit.
Indeed, all parameters in the rotational Hamiltonian (\ref{rotmass})  have definite $N_c$ dependence:
\begin{equation}
M^{(K)}_{\text{sol}},\, I_{1,2} \sim \mathcal{O}(N_c), ~~a_K \sim \mathcal{O}(1)\, .
\end{equation}
Therefore rotational excitations of the soliton (diquark excitations in the quark language)
have energy splittings which are parametrically small $1/I_{1,2} \sim 1/N_c$.

On the other hand, $\lambda$ modes correspond, in principle, to a color confining 
soliton-$Q$ (or diquark-$Q$) interaction that
could be taken into account by solving a Schr{\"o}dinger equation with reduced mass $\mu$. In a heavy
quark limit  $\mu \sim N_c$. Excitation energy $\Delta_{\lambda}E$ dependence on $\mu$ is closely
related to a potential used and is linear in $\mu$ for the Coulomb potential. For the logarithmic potential it does not depend
on $\mu$, and for the linear potential it is proportional to $\mu^{-1/3}$. In all these cases mass splittings corresponding to the
soliton $\rho$ modes are parametrically
smaller than the energy splittings of the $\lambda$ modes and are therefore dominant. This should be contrasted with the quark models
where typically $\lambda$ modes are dominant ({\em e.g.}~\cite{Bijker:2020tns,Cheng:2015naa}).

The rotational Hamiltonian (\ref{rotmass}) has to be supplemented by the SU(3)
symmetry-breaking {collective} Hamiltonian and the hyperfine splitting
Hamiltonian, which we parametrize as follows~\cite{Yang:2016qdz}:
\begin{equation}
H_{\mathrm{hf}}=\frac{2}{3}\frac{\kappa}{m_{Q}}\bm{J}\cdot\bm{J}_{Q}
\label{eq:ssinter}%
\end{equation}
where $\kappa$ is flavor-indepenent free parameter that may depend on the
SU(3) representation and on the soliton grand spin $K$. The operators
${\bm{J}}$ and ${\bm{J}}_{Q}$ represent the spin operators for the soliton and
the heavy quark, respectively. As discussed in Sec.~\ref{ssec:Skyrme}, such
interaction term can be derived from the bound-state approach to the Skyrme model.

The symmetry breaking Hamiltonian, which can be found in
Ref.~\cite{Kim:2017jpx}, is given in terms of the SU(3) $D$ functions,
hypercharge $Y$, ${\bm T}^{\prime}$ isospin operators and grand spin ${\bm K}$
operator. It is parametrized by four constants: $\alpha$, $\beta$, $\gamma$
and $\delta$. When this Hamiltonian is sandwiched between the
{collective} wave functions, one obtains that mass splittings in
antitriplet and sextet are proportional to the hypercharge $Y$, with
coefficients given as some combinations of the parameters $\alpha,\ldots,\,
\delta$. We refer the reader to Refs.~\cite{Kim:2017jpx,Yang:2016qdz} for
their explicit forms.

It is convenient to introduce the following quantities for the ground state
baryons
\begin{align}
\mathcal{M}_{Q\,\overline{\mathbf{3}}} &  =m_{Q}+M_{\mathrm{sol}}+\frac{1}{2
I_{2}}\, ,\nonumber\\
\mathcal{M}_{Q\, \mathbf{6}} &  =\mathcal{M}_{Q \, \overline{\mathbf{3}}%
}+\frac{1}{I_{1}}\label{eq:MQ}%
\end{align}
and for $K=1$ excited multiplets
\begin{align}
\mathcal{M}^{\prime}_{Q\,\overline{\mathbf{3}}} &  =m_{Q}+M^{\prime
}_{\mathrm{sol}}+\frac{1}{2 I^{\prime}_{2}}+\frac{a_{1}^{2}}{I^{\prime}_{1}}\,
,\nonumber\\
\mathcal{M}^{\prime}_{Q\, \mathbf{6}} &  =\mathcal{M}^{\prime}_{Q \,
\overline{\mathbf{3}}}+\frac{1}{I^{\prime}_{1}}\, .\label{eq:MQprime}%
\end{align}
Here primes indicate that both the soliton mass $M_{\mathrm{sol}}$ and the moments of
inertia $I_{1,2}$ calculated for the excited configuration are numerically
different from the ones calculated for the ground state. Also the soliton mass
and moments of inertia for the ground state heavy baryons may differ from the
ones in the light sector. These differences are in principle of the order of
$1/N_{c}$ and therefore are negligible for large $N_{c}$, but might be
important for the real world $N_{c}=3$ phenomenology.

With the help of Eqs.~(\ref{eq:MQ}) we can write concise formulas for the
ground states,
\begin{align}
M_{Q\, Y}^{\overline{\mathbf{3}}}  &  =\mathcal{M}_{Q \,\overline{\mathbf{3}}%
}+ \delta_{\overline{\mathbf{3}}} Y \, ,\nonumber\\
M_{Q\, Y}^{\mathbf{6}}  &  =\mathcal{M}_{Q \,\mathbf{6}}+ \delta_{\mathbf{6}}
Y+\frac{\kappa}{m_{Q}}\left\{
\begin{array}
[c]{ccc}%
-2/3 & \text{for} & S=1/2\\
\, & \, & \,\\
+1/3 & \text{for} & S=3/2
\end{array}
\right.  \label{eq:6massesg}%
\end{align}
and using Eqs.~(\ref{eq:MQprime}) for negative-parity excited states
\begin{widetext}
\begin{align}
{M^{\prime}}_{Q\, Y}^{\overline{\mathbf{3}}}  &  =\mathcal{M}_{Q
\,\overline{\mathbf{3}}}^{\prime}+ \delta_{\overline{\mathbf{3}}}
Y+\frac{\kappa^{\prime}}{m_{Q}}\left\{
\begin{array}
[c]{ccc}%
-2/3 & \text{for} & S=1/2\\
\, & \, & \,\\
+1/3 & \text{for} & S=3/2
\end{array}
\right. \, ,\nonumber\\
~ & ~ \nonumber \\
{M^{\prime}}_{Q\, Y}^{\mathbf{6}\, J=0}  &  =\mathcal{M}_{\mathbf{6}\,
Q}^{\prime}-2\frac{a_{1}}{I^{\prime}_{1}}+\left(  \delta_{\mathbf{6}}-\frac
{3}{10}\delta\right)  Y,\nonumber\\
{M^{\prime}}_{Q \,Y}^{\mathbf{6}\, J=1}  &  =\mathcal{M}_{\mathbf{6}\,
Q}^{\prime}-\frac{a_{1}}{I^{\prime}_{1}}+\left(  \delta_{\mathbf{6}}-\frac
{3}{20}\delta\right)  Y+\frac{\kappa^{\prime}}{m_{Q}}\left\{
\begin{array}
[c]{ccc}%
-2/3 & \text{for} & S=1/2\\
\, & \, & \,\\
+1/3 & \text{for} & S=3/2
\end{array}
\right. \, , \nonumber\\
{M^{\prime}}_{Q \, Y}^{\mathbf{6}\, J=2}  &  =\mathcal{M}_{\mathbf{6}\,
Q}^{\prime}+\frac{a_{1}}{I^{\prime}_{1}}+\left(  \delta_{\mathbf{6}}+\frac
{3}{20}\delta\right)  Y+\frac{\kappa^{\prime}}{m_{Q}}\left\{
\begin{array}
[c]{ccc}%
-1 & \text{for} & S=3/2\\
\, & \, & \,\\
+2/3 & \text{for} & S=5/2
\end{array}
\right. \, . \label{eq:6massese}%
\end{align}
\end{widetext}
Here $\delta_{\overline{\mathbf{3}}}$ and $\delta_{\mathbf{6}}$ are effective breaking
parameters constructed from $\alpha$, $\beta$ and $\gamma$, and $\delta$ is a
new parameter that enters the splittings of the excited sextets only \cite{Yang:2016qdz}.
We see that sextet mass splittings depend on $J$ and, therefore, care must be taken when
when applying the Eckart-Wigner theorem to these multiplets (see {\em e.g.} \cite{Bijker:2020tns}).
We will come back to this point later.

An immediate consequence of Eqs.~(\ref{eq:6massesg}) and (\ref{eq:6massese})
for antitriplets is the equality of splittings in ground and excited states,
already mentioned in Sec.~\ref{sec:exp}. This property goes beyond the usual
Eckart-Wigner type SU(3) relations and is a consequence of the {hedgehog}
symmetry. From Tables~\ref{tab:3bar_c} and \ref{tab:3bar_b}, and from
Eq.~(\ref{eq:Xibpredictions}) we can estimate average excited antitriplet
masses (in MeV/$c^{2}$),
\begin{equation}
\mathcal{M}_{\overline{\mathbf{3}}\, c}^{\prime}=2745, ~~~\mathcal{M}%
_{\overline{\mathbf{3}}\, b}^{\prime}=6034\, ,\label{eq:Mav3bar}%
\end{equation}
which give heavy quark mass difference $m_{b}-m_{c}=3289$~MeV/$c^{2}$,
slightly below (\ref{eq:bcdiff}).

\section{Phenomenology}

\label{sect:pheno}

\subsection{Ground state heavy baryons}

\label{ssec:pheno_ground}

In this subsection we briefly recapitulate phenomenology of the ground state
heavy baryons based on mass formulas (\ref{eq:6massesg}), discussed initially
in \cite{Yang:2016qdz}. As already mentioned in Sec.~\ref{sec:exp} both
antitriplet and and sextet splittings have little dependence on the heavy
quark mass [see Eqs.~(\ref{eq:delta3barpheno}], (\ref{eq:6cpheno}) and
(\ref{eq:6bpheno})). Throughout this paper in the following analysis we shall
use an average value
\begin{equation}
\delta_{\bm6}=-118~\mathrm{MeV}/c^{2}\, .%
\end{equation}

Furthermore the hyperfine splittings for the ground state charm sextet give
(see Table~\ref{tab:6_c})
\begin{equation}
\frac{\kappa}{m_{c}} = 64 \div71~\mathrm{MeV}/c^{2} \, .\label{eq:kappac}%
\end{equation}
This value of the hyperfine splitting resulted in assigning the two excited
$\Omega_{c}$ states~\cite{LHCb:2017uwr} of masses 3050~MeV/$c^{2}$ and 3119~MeV/$c^{2}$
(see Table~\ref{tab:6_c}) as members of $\overline{\bm{15}}$ exotic SU(3)
multiplet \cite{Kim:2017jpx}. Subsequent calculation of the decay widths
\cite{Kim:2017khv}, which for exotic representation $\overline{\bm{15}}$ are
expected to be small in the large $N_{c}$ limit~\cite{Praszalowicz:2018upb},
strengthened this assignment.

In the beauty sector (see Table~\ref{tab:6_b})
\begin{equation}
\frac{\kappa}{m_{b}} \approx20~\mathrm{MeV}/c^{2} \, .\label{eq:kappab}%
\end{equation}
in agreement with simple scaling of the hyperfine splitting with the quark masses.

From mass centres of antitriplet and and sextet multiplets we can calculate
the difference of the heavy quark masses. For ${\overline{\bm3}}$ (from
Tables~\ref{tab:3bar_c} and \ref{tab:3bar_b}) we have (in what follows, to
simplify notation, we will use particle symbols for their masses):%
\begin{equation}
m_{b}-m_{c}=\frac{1}{3}\left( \Lambda^{1/2^{+}}_{Q}+2\Xi^{1/2^{+}}_{Q}\right)
\Big|_{b-c}=3328~\mathrm{MeV}/c^{2}\, .\label{eq:diffbc1}%
\end{equation}
For ${\bm 6}$ we have to average over spin. Using
Tables~\ref{tab:6_c} and \ref{tab:6_b} one gets,
\begin{widetext}
\begin{align}
m_{b}-m_{c} & =\frac{1}{6} \left( 3\frac{\Sigma_{Q}^{1/2^{+}}+2 \Sigma
_{Q}^{3/2^{+}}}{3} +2\frac{\Xi_{Q}^{1/2^{+}}+2 \Xi_{Q}^{3/2^{+}}}{3}
+\frac{\Omega_{Q}^{1/2^{+}}+2 \Omega_{Q}^{3/2^{+}}}{3}\right) \Bigg|_{b-c}%
=3327~\mathrm{MeV}/c^{2}\, .\label{eq:diffbc2}%
\end{align}
We see perfect agreement of both estimates of the difference of heavy quark
masses with (\ref{eq:bcdiff}) (see Ref.~\cite{Praszalowicz:2019lje} the discussion of the
analogous relation for heavy mesons).
\end{widetext}

\subsection{Excited antitriplets}

\label{ssec:pheno_excited3bar}

As already mentioned splitting parameter $\delta_{\bar{{\boldsymbol{3}}}}$
[see first equation in (\ref{eq:6massese})] is identical to the one of the
ground state, both for charm and beauty with accuracy better than $15\%$.

Furthermore, for charm we can extract the hyperfine splitting parameter from
the following mass differences
\begin{align}
\frac{\kappa^{\prime}}{m_{c}} & =\Lambda_{c}^{3/2^{-}}(2628)-\Lambda
_{c}^{1/2^{-}}(2592)=36~\text{ MeV}/c^{2} \, ,\nonumber\\
\frac{\kappa^{\prime}}{m_{c}} & =\Xi_{c}^{3/2^{-}}(2818)-\Xi_{c}^{1/2^{-}%
}(2793)=25~\text{ MeV}/c^{2} \, .\label{eq:kappacprime}%
\end{align}
This relatively big difference ($\sim17\%$) between the two estimates may be
attributed to an imperfect heavy quark symmetry for charm.

For beauty we have
\begin{equation}
\frac{\kappa^{\prime}}{m_{b}}=\Lambda_{b}^{3/2^{-}}(5920)-\Lambda_{b}%
^{1/2^{-}}(5912)=8\text{ MeV}/c^{2} \, .\label{eq:bhf}%
\end{equation}
This is in good agreement with the charm estimate if we recall that
$m_{c}/m_{b}\simeq0.3$.

From spin averaged masses of $\Lambda_{Q}$ baryons\footnote{Unlike in the case
of the ground state antitriplets, $\Xi_{b}^{1/2^{-}}$ has not been measured.}%
\begin{equation}
\frac{1}{3}\left(  \Lambda_{b}^{1/2^{-}}+2\Lambda_{b}^{3/2^{-}}\right)
=\mathcal{M}_{\mathbf{\bar{3}}}^{\prime}+\delta_{\boldsymbol{\bar{3}}}Y
\end{equation}
we can calculate
\begin{equation}
m_{b}-m_{c}=\frac{1}{3}\left(  \Lambda_{Q}^{1/2^{-}}+2\Lambda_{Q}^{3/2^{-}%
}\right) \Big|_{b-c} =3302\;\text{MeV}/c^{2}%
\end{equation}
in agreement with our previous estimates (\ref{eq:diffbc1}) and
(\ref{eq:diffbc2}).

Finally, let us observe that the model {predicts} with no free parameters
masses of $\Xi_{b}^{1/2^{-}}$ and $\Xi_{b}^{3/2^{-}}$, as already shown in
Eq.~(\ref{eq:Xibpredictions}).

\subsection{Excited sextets}
\label{subs:6prim}

There are three excited sextets characterized by the soliton spin $J=0,1,2$.
The spin $S$ of the heavy baryon emerges from the heavy quark coupling with
the soliton spin. So, we have five sextets:
$(J=0,S=1/2),\;(J=1,S=1/2),\;(J=1,S=3/2),\;(J=2,S=3/2)$, and$\;(J=2,S=5/2)$ all
of parity $P=-$. Therefore we expect 5 isospin multiplets of negative-parity
$\Sigma_{Q}$'s and $\Xi_{Q}$'s (not to mention positive parity radial
excitations of the ground state multiplets). As seen from
Tables~\ref{tab:unknown_c} and \ref{tab:unknown_b} we have only one
$\Sigma_{Q}$ isospin multiplet candidate both for charm and for beauty.
The situation is somewhat better in the $\Xi_{Q}$ case; however, here we have to
remember that $\Xi_{Q}$ candidates may belong both to $\overline{\bm 3}$ and
$\bm 6$ SU(3) multiplets. The only unambiguous sextet candidates are excited
$\Omega_{c}^{0}$ states (see Table~\ref{tab:6_c}) -- or at least some of them
-- reported by LHCb \cite{LHCb:2021ptx}. Before we recall the possible assignments
of the of $\Omega_{c}^{0}$'s proposed in Ref.~\cite{Kim:2017jpx}, let us
discuss some general features of the sextet spectra.

It seems that that the predictive power of the sextet formulas of
Eqs.~(\ref{eq:6massese}) is rather weak, since we have two new parameters:
$\delta$ that makes hypercharge splittings in the excited sextet different
from the ones of the ground state, and $a_{1}$ which is responsible for
splittings of different $J$ multiplets. However, as has been observed in
Ref.~\cite{Kim:2017jpx}, splittings of different $J$ multiplets before
hyperfine splitting ({i.e.} for $\kappa^{\prime}=0$)
\begin{align}
\Delta_{1}(Y)&=\left.  \left(  {M^{\prime}}_{Q\,Y}^{{\bm 6}J=1}-{M^{\prime}%
}_{Q\,Y}^{{\bm 6}J=0}\right)  \right\vert _{\kappa^{\prime}=0},\notag \\
\Delta
_{2}(Y)&=\left.  \left(  {M^{\prime}}_{Q\,Y}^{{\bm 6}J=2}- {M^{\prime}}%
_{Q\,Y}^{{\bm 6}J=1}\right)  \right\vert _{\kappa^{\prime}=0}%
\end{align}
do not depend on $Q$ and read as follows
\begin{equation}
\Delta_{1}(Y)=\frac{a_{1}}{I_{1}}+\frac{3}{20}\delta\,Y
\end{equation}
and
\begin{equation}
\Delta_{2}(Y)=2\Delta_{1}(Y) \, ,\label{eqD1vsD2}%
\end{equation}
which significantly constrains sextet spectra.


As said previously, the only unambiguous candidates for the excited sextets
are five LHCb \cite{LHCb:2021ptx}  $\Omega_{c}^{0}$'s. However, it is
impossible to assign all of them to these five sextets~\cite{Kim:2017jpx}  due
to the values of the hyperfine splittings and the constraint (\ref{eqD1vsD2}).
Moreover, two  of the LHCb $\Omega_{c}^{0}$'s are very narrow, around 1 MeV,
while the remaining three have widths  ranging from 3.4 MeV to 8.7 MeV. Such
spread of the decay widths would greatly violate SU(3) relations between the
decay constants.

Therefore it has been proposed in Ref.~\cite{Kim:2017jpx} to interpret two
narrow $\Omega$ states, namely $\Omega_{c}^{0}(3050)$ and $\Omega_{c}^{0}(3119)$,
as the hyperfine split members of the exotic $\overline{\bm{15}}$. This
assignment has been motivated by the fact\footnote{Note that the ground state
sextet and exotic $\overline{\bm{15}}$ belong to the same rotational band, and
therefore should have approximately the same value of $\kappa$.} that their
hyperfine splitting is equal to the one of the ground state sextet
(\ref{eq:kappac}).

Leaving aside the interpretation of $\Omega_{c}^{0}(3050)$ and $\Omega_{c}%
^{0}(3119)$, we are left with three remaining sates $\Omega_{c}^{0}%
(3000),~\Omega_{c}^{0}(3065)$, and $\Omega_{c}^{0}(3090)$, which have been
interpreted as members of $J=0$ and $J=1$ sextets~\cite{Kim:2017jpx}. Indeed,
if $\Omega_{c}^{0}(3065)$ and $\Omega_{c}^{0}(3090)$ are hyperfine split
members of $J=1$ sextet, then their hyperfine splitting should be equal to the
one of the excited antitriplet (\ref{eq:kappacprime}), which is indeed the
case. From this assignment one obtains,
\begin{equation}
\Delta_{1}(Y=-4/3)=\frac{a_{1}}{I_{1}}-\frac{1}{5}\delta\,=82\;\text{MeV}%
/c^{2} \, . \label{eq:Delta1}%
\end{equation}

In view of relation (\ref{eqD1vsD2}) we should have two other $\Omega_{c}^{0}$
states approximately 164~MeV/$c^{2}$ higher. Indeed, in
Ref.~\cite{Kim:2017jpx} masses of two hyperfine split members of $J=2$ sextet
have been estimated to be 3222 MeV/$c^{2}$ and 3262 MeV/$c^{2}$. In this scenario these
states have masses above the $\Xi+D$ threshold at 3185 MeV/$c^{2}$, {i.e.} they
can have rather large widths and may be not clearly seen in the LHCb
data\footnote{Note that LHCb sees some wide bumps at 3188~MeV/$c^{2}$ and higher.}.


Recent report of the LHCb Collaboration~\cite{LHCb-talk} on $\Xi_{b}%
^{-}(6327)$ and $\Xi_{b}^{-}(6333)$ allows for narrowing the model parameter
space. Indeed, these -- together with $\Xi_{b}(6227)$ and possibly $\Sigma
_{b}(6097)$ -- are the only inputs in the $b$ sector, which can be used. Four
$\Omega_{b}^{-}$ states are only one star resonances and therefore cannot give
reliable information to constrain the model. On the contrary $\Xi_{b}%
^{-}(6327)$ and $\Xi_{b}^{-}(6333)$ fit very well a hypothesis that they are
hyperfine split members of $J=1$ excited sextet. Hyperfine splitting is in
this case $\sim5.5$~MeV/$c^{2}$ in accordance with an expectation from the $b$
antitriplet of 8~MeV/$c^{2}$ (\ref{eq:bhf}). Note that hyperfine splittings
have rather large model uncertainty, see estimates for charm
(\ref{eq:kappacprime}).

Therefore we propose the following assignments,
\begin{align}
\Sigma_{b}(6097)  &  =\Sigma_{b}^{1/2^{-}}(\boldsymbol{6}^{\prime},J=0) \,
,\nonumber\\
\Xi_{b}(6327)  &  =\Xi_{b}^{1/2^{-}}(\boldsymbol{6}^{\prime},J=1)\,
,\nonumber\\
\Xi_{b}(6333)  &  =\Xi_{b}^{3/2^{-}}(\boldsymbol{6}^{\prime}%
,J=1)\,\label{eq:binputs}%
\end{align}
where we have also used the LHCb~\cite{LHCb:2018haf} $\Sigma_{b}(6097)$. Other
assignments of $\Sigma_{b}(6097)$ give much worse fits.
One might try to assign $\Xi_{b}(6327)$  and $\Xi_{b}(6333)$ to $(\boldsymbol{6}^{\prime}%
,J=2)$, however for $J=2$ 
the hyperfine splitting is expected to be $\sim 13.3$~MeV, see Eqs.~(\ref{eq:6massese}),
rather than 8~MeV  for $J=1$ (vs. experimental 5.5~MeV). Therefore we use 
(\ref{eq:binputs}) in the following fits.

In Fit 0 we fix the unknown model parameters, namely $\delta$ and
$a_{1}/I^{\prime}_{1}$ [or equivalently $\Delta_{1}(-4/3)$, see
Eq.~(\ref{eq:Delta1})] using all masses in Eq.~(\ref{eq:binputs}) as inputs.
The results of this fit are shown in Table~\ref{tab:test0}. The resulting
$\Delta_{1}(-4/3)=98$~MeV/$c^{2}$ is a bit higher than the one obtained from
the charm sector (\ref{eq:Delta1}). In order to make contact with the charm
spectrum in the three following fits we fix $\Delta_{1}(-4/3)=82$~MeV/$c^{2}$
[which follows from three $\Omega^{0}_{c}$ states used as an input (\ref{eq:Delta1})] and use
different combinations of two inputs from Eqs.~(\ref{eq:binputs}). The results
are presented in Tabs.~\ref{tab:test1}~--~\ref{tab:test3}, where also the
resulting charm spectrum is shown. Entries in bold face have been used as
inputs, underlined entries denote masses that can be attributed to some of the
baryons of the unknown SU(3) assignment listed in Tables~\ref{tab:unknown_c}
and \ref{tab:unknown_b}: solid line if mass difference is of the order of
12~MeV/$c^{2}$ or less, dashed line if the mass difference is
15~~MeV/$c^{2}$--~20~MeV/$c^{2}$.


\begin{table}[h]
\centering
\begin{tabular}
[c]{|cc|ccc|}\hline
$J$ & $S$ & $\Sigma_{b}$ & $\Xi_{b} $ & $\Omega_{b}$\\\hline
0 & 1/2 & \textbf{6097} & \textbf{6227} & 6357\\\hline
1 & $%
\begin{array}
[c]{c}%
1/2\\
3/2
\end{array}
$ & $%
\begin{array}
[c]{c}%
6203\\
6209
\end{array}
$ & $%
\begin{array}
[c]{c}%
\bm{6327}\\
\bm{6333}
\end{array}
$ & $%
\begin{array}
[c]{c}%
6451\\
6557
\end{array}
$\\\hline
2 & $%
\begin{array}
[c]{c}%
3/2\\
5/2
\end{array}
$ & $%
\begin{array}
[c]{c}%
6422\\
6431
\end{array}
$ & $%
\begin{array}
[c]{c}%
6534\\
6543
\end{array}
$ & $%
\begin{array}
[c]{c}%
6646\\
6655
\end{array}
$\\\hline
\end{tabular}
\caption{Fit 0: $\delta=40$ and $\Delta_{1}(-4/3)=98$. Inputs are in bold
face. All masses are in MeV/$c^{2}$.}%
\label{tab:test0}%
\end{table}


\begin{table}[h]
\centering
\begin{tabular}
[c]{|cc|ccc||ccc|}\hline
$J$ & $S$ & $\Sigma_{b}$ & $\Xi_{b} $ & $\Omega_{b}$ & $\Sigma_{c}$ & $\Xi_{c}
$ & $\Omega_{c}$\\\hline
0 & 1/2 & \textbf{6097} & \underline{6238} & 6378 & 2719 & 2859 &
\textbf{3000}\\\hline
1 & $%
\begin{array}
[c]{c}%
1/2\\
3/2
\end{array}
$ & $%
\begin{array}
[c]{c}%
6198\\
6204
\end{array}
$ & $%
\begin{array}
[c]{c}%
\bm{6327}\\
\bm{6333}
\end{array}
$ & $%
\begin{array}
[c]{c}%
6457\\
6462
\end{array}
$ & $%
\begin{array}
[c]{c}%
\underline{2807}\\
2831
\end{array}
$ & $%
\begin{array}
[c]{c}%
\underline{2937}\\
\underline{2961}%
\end{array}
$ & $%
\begin{array}
[c]{c}%
\bm{3066}\\
\bm{3090}
\end{array}
$\\\hline
2 & $%
\begin{array}
[c]{c}%
3/2\\
5/2
\end{array}
$ & $%
\begin{array}
[c]{c}%
6406\\
6415
\end{array}
$ & $%
\begin{array}
[c]{c}%
6512\\
6521
\end{array}
$ & $%
\begin{array}
[c]{c}%
6619\\
6628
\end{array}
$ & $%
\begin{array}
[c]{c}%
3009\\
3049
\end{array}
$ & $%
\begin{array}
[c]{c}%
\underline{3115}\\
3155
\end{array}
$ & $%
\begin{array}
[c]{c}%
3222\\
3262
\end{array}
$\\\hline
\end{tabular}
\caption{Fit 1, $\delta=76$. Inputs are in bold face. Underlined entries
can be attributed to some of the known baryons (see text). All masses are in
MeV/$c^{2}$.}%
\label{tab:test1}%
\end{table}

\begin{table}[h]
\centering
\begin{tabular}
[c]{|cc|ccc||ccc|}\hline
$J$ & $S$ & $\Sigma_{b}$ & $\Xi_{b} $ & $\Omega_{b}$ & $\Sigma_{c}$ & $\Xi_{c}
$ & $\Omega_{c}$\\\hline
0 & 1/2 & 6065 & \textbf{6227} & 6389 & 2676 & 2838 & \textbf{3000} \\\hline
1 & $%
\begin{array}
[c]{c}%
1/2\\
3/2
\end{array}
$ & $%
\begin{array}
[c]{c}%
6187\\
6193
\end{array}
$ & $%
\begin{array}
[c]{c}%
\bm{6327}\\
\bm{6333}
\end{array}
$ & $%
\begin{array}
[c]{c}%
6467\\
6473
\end{array}
$ & $%
\begin{array}
[c]{c}%
2786\\
\underline{2810}%
\end{array}
$ & $%
\begin{array}
[c]{c}%
\underline{2926}\\
\dashuline{~2950~}
\end{array}
$ & $%
\begin{array}
[c]{c}%
\bm{3066}\\
\bm{3090}
\end{array}
$\\\hline
2 & $%
\begin{array}
[c]{c}%
3/2\\
5/2
\end{array}
$ & $%
\begin{array}
[c]{c}%
6438\\
6447
\end{array}
$ & $%
\begin{array}
[c]{c}%
6534\\
6543
\end{array}
$ & $%
\begin{array}
[c]{c}%
6630\\
6639
\end{array}
$ & $%
\begin{array}
[c]{c}%
3030\\
3070
\end{array}
$ & $%
\begin{array}
[c]{c}%
\underline{3126}\\
3166
\end{array}
$ & $%
\begin{array}
[c]{c}%
3222\\
3262
\end{array}
$\\\hline
\end{tabular}
\caption{Fit 2, $\delta=147$. Inputs are in bold face. Underlined entries
can be attributed to some of the known baryons (see text). All masses are in
MeV/$c^{2}$.}%
\label{tab:test2}%
\end{table}


\begin{table}[h]
\centering
\begin{tabular}
[c]{|cc|ccc||ccc|}\hline
$J$ & $S$ & $\Sigma_{b}$ & $\Xi_{b} $ & $\Omega_{b}$ & $\Sigma_{c}$ & $\Xi_{c}
$ & $\Omega_{c}$\\\hline
0 & 1/2 & \textbf{6097} & \textbf{6227} & 6357 & 2740 & 2870 & \textbf{3000}%
\\\hline
1 & $%
\begin{array}
[c]{c}%
1/2\\
3/2
\end{array}
$ & $%
\begin{array}
[c]{c}%
6188\\
6193
\end{array}
$ & $%
\begin{array}
[c]{c}%
\dashuline{~6312~}\\
\dashuline{~6317~}
\end{array}
$ & $%
\begin{array}
[c]{c}%
6436\\
6441
\end{array}
$ & $%
\begin{array}
[c]{c}%
\dashuline{~2818}\\
2842
\end{array}
$ & $%
\begin{array}
[c]{c}%
\underline{2942}\\
\underline{2966}%
\end{array}
$ & $%
\begin{array}
[c]{c}%
\bm{3066}\\
\bm{3090}
\end{array}
$ \\\hline
2 & $%
\begin{array}
[c]{c}%
3/2\\
5/2
\end{array}
$ & $%
\begin{array}
[c]{c}%
6374\\
6382
\end{array}
$ & $%
\begin{array}
[c]{c}%
6486\\
6494
\end{array}
$ & $%
\begin{array}
[c]{c}%
6598\\
6606
\end{array}
$ & $%
\begin{array}
[c]{c}%
2998\\
3038
\end{array}
$ & $%
\begin{array}
[c]{c}%
\dashuline{~3110~}\\
3150
\end{array}
$ & $%
\begin{array}
[c]{c}%
3222\\
3262
\end{array}
$\\\hline
\end{tabular}
\caption{Fit 3, $\delta=40$. Inputs are in bold face. Underlined entries can
be attributed to some of the known baryons (see text). All masses are in
MeV/$c^{2}$.}%
\label{tab:test3}%
\end{table}

\newpage

We see that although fits 1 -- 3 give rather different values of $\delta$,
they confirm, with possible exception of fit 2 that misses $\Sigma_{b}(6097)$
by 32~MeV/$c^{2}$, the hypothesis of Eq.~(\ref{eq:binputs}). Moreover, they
suggest clear interpretation of four charm states from
Table~\ref{tab:unknown_c}:
\begin{align}
\Sigma_{c}(2800) &  = \Sigma_{c}^{1/2^{-}}(\bm{6}^{\prime},J=1) \,
,\nonumber\\
\Xi_{c}(2940)  &  = \Xi_{c}^{1/2^{-}}(\bm{6}^{\prime},J=1) \, ,\nonumber\\
\Xi_{c}(2966)  &  = \Xi_{c}^{3/2^{-}}(\bm{6}^{\prime},J=1) \, ,\nonumber\\
\Xi_{c}(3123)  &  = \Xi_{c}^{3/2^{-}}(\bm{6}^{\prime},J=2) \,
.\label{eq:chypoth}%
\end{align}
Interestingly, a three star $\Xi_{c}(3056)$ cannot be interpreted as a member
of negative-parity $\bm{6}$, while a two star $\Xi_{c}(2940)$ seems to be a spin
partner of a three star $\Xi_{c}(2966)$, and one star $\Xi_{c}(3123)$ is
interpreted as $J=2$ sextet excitation. Moreover, there are no experimental candidates for
the lowest $\bm{6}$ multiplet of $J=0$.

With these assignments we can calculate average sextet masses both for charm
and bottom baryons (in MeV/$c^{2}$)\footnote{We use Fit 1.},
\begin{equation}
\mathcal{M}_{\mathbf{6}\,c}^{\prime}=3007,~~~\mathcal{M}_{\mathbf{6}%
\,b}^{\prime}=6385\,\label{eq:Mav6}%
\end{equation}
which results in $m_{b}-m_{c}=3378$~MeV/$c^{2}$, {i.e.} $\sim 70$~MeV/$c^{2}$
above (\ref{eq:bcdiff}). These differences of $m_{b}-m_{c}$ extraction from
average masses of excited multiplets (\ref{eq:Mav3bar}) and (\ref{eq:Mav6})
together with perfect agreement of $m_{b}-m_{c}$ extraction from the ground
sates (\ref{eq:diffbc1}) and (\ref{eq:diffbc2}) suggest that we miss some
contributions to the overall masses. This can be further exemplified when one
realizes that the sextet masses are in principle calculable in the present
approach, see Eqs.~(\ref{eq:MQ}) and (\ref{eq:MQprime}). Indeed, from the
ground state antitriplet and sextet we obtain $1/I_{1}=172$~MeV/$c^{2}$ and
170~MeV/$c^{2}$ for charm and beauty, respectively (see
Ref.~\cite{Yang:2016qdz} for discussion). In the excited sector we get
$1/{I_{1}}^{\prime}=262$~MeV/$c^{2}$ and 351~MeV/$c^{2}$ for charm and beauty,
respectively. Two remarks are here in order. We expect some numerical
difference between $1/I_{1}$ and $1/{I_{1}}^{\prime}$, but 90~MeV/$c^{2}$ is
larger than anticipated. What is more troubling, is the large difference
between charm and beauty estimates. A practical solution would be to lower the
average sextet mass for beauty, which would also improve $m_{b}-m_{c}$. This,
however, would require to associate two LHCb states $\Xi_{b}(6327)$ and
$\Xi_{b}(6333)$ with $J=2$ sextet, rather with $J=1$. However, the expected
hyperfine splitting in $J=2$ sextet is almost twice larger than for $J=1$, and
this would be completely incompatible with the charm sector.
On the
other hand even if we took $1/{I_{1}}^{\prime}=1/I_{1}$ and applied this to
the excited sextets, we would underestimate $\mathcal{M}_{\mathbf{6}%
\,Q}^{\prime}$ only by $\sim3\%$ both for charm and beauty.

\section{Discussion}
\label{sect:disc}

Observation of five $\Omega_c^0$ states by the  LHCb Collaboration \cite{LHCb:2017uwr}
 implies that we should expect similar states in the beauty sector.
Indeed, the replacement of the $c$-quark by the $b$-quark leads only to the overall mass shift of 
approximately 3330 MeV/$c^2$
[see Eq.~(\ref{eq:bcdiff})] and to the rescaling of the hyperfine splittings. 
On general grounds the mass splittings inside the excited multiplets
and splittings between various multiplets should not change under the replacement of the $c$-quark 
by the $b$-quark. So, we  expect at least five narrow 
$\Omega_b$'s populating the mass interval 6350~MeV/$c^2$ -- 6650~MeV/$c^2$ (see Tables \ref{tab:test0}--\ref{tab:test3}) 
and the same number of the corresponding $\Xi_Q$ and $\Sigma_Q$ states in both
sectors.

However, we see from the review of present experimental data of Sec.~\ref{sec:exp} that only a few
candidates have been found so far. Why do we not see all the states required by this simple argument?
Possible explanations are:
\begin{itemize}
\item Some of the $\boldsymbol{6}'$ states  have small couplings to the ground state
$\overline{\boldsymbol{3}}$. This is indeed the case, as illustrated in Table~\ref{tab:dcouplings}.
\item For some reason the observed peak is not one resonance but several ones almost degenerated in 
mass. Such possibility is admitted {\em e.g.} by the LHCb Collaboration~\cite{LHCb:2018haf}.
\item It could be that some of expected excited sextet states are very wide and therefore are not seen 
experimentally. This seems to be the case for the $J=2$ sextet, where the allowed decays are in $D$
wave (see Table~\ref{tab:dcouplings}).
\item Some of the excited states are very narrow and the experimental resolution is not good enough
to detect them.
\end{itemize}
In the following we shall review from this point of view  some recent experimental searches of the 
 $\Sigma_Q$ and $\Xi_Q$
excited states.

\begin{table}
\centering
\begin{tabular}
[c]{|c|ccc|}%
\hline
\;  & $ \overline{\boldsymbol{3}}(J^{P}=0^{+}) $  &  $ \boldsymbol{6}(J^{P}=1^{+}) $ &
$ \overline{\boldsymbol{3}}^{\prime}(J^{P}=1^{-}) $\\
\hline
$\overline{\boldsymbol{3}}^{\prime}(J^{P}=1^{-}) $&$ 1/m_{Q} $&$ S $& $-$\\
$\boldsymbol{6}^{\prime}(J^{P}=0^{-}) $& $S $& $1/m_{Q} $&$ P$\\
$\boldsymbol{6}^{\prime}(J^{P}=1^{-}) $&  $1/m_{Q} $ & $ S $ & $ P$\\
$\boldsymbol{6}^{\prime}(J^{P}=2^{-}) $&$ D$ & $D$ &$ P$\\
\hline
\end{tabular}
\caption{Type of couplings for the decays of excited heavy baryons to ground state multiplets
and excited $\overline{\boldsymbol{3}}^{\prime}$ plus a pseudo scalar meson. Note that in the heavy
quark limit  $c$- and $b\,$-quarks act as spectators when the soliton decays. If the decay is forbidden by the
angular momentum conservation, then the heavy quark spin flip may provide the missing angular momentum,
but there is a penalty for the spin flip of $1/m_Q$ in the decay amplitude. }
\label{tab:dcouplings}
\end{table}

\subsection{Where are the missing $\Sigma_Q$ states?}

Let us start with a $b$ sector and the LHCb observation of $\Sigma_b(6097)$ \cite{LHCb:2018haf} in the 
the distribution of  $\Lambda^0_b$ and $\pi^{\pm}$ mass over the range between 5760~MeV/$c^2$ -- 6360~MeV/$c^2$.
One can clearly see (Fig. 2 in Ref.~\cite{LHCb:2018haf}) two small peaks corresponding to the ground state
sextet; $\Sigma_b(5813)$ and $\Sigma_b(5833)$, and the large peak of $\Sigma_b(6097)$.
With our interpretation (\ref{eq:binputs})
in this energy range at approximately
 6200~MeV/$c^2$ we expect two hyperfine split resonances
corresponding to $(\boldsymbol{6}',J=1)$ multiplet\footnote{This energy corresponds to 
$Q \sim 440$~MeV/$c^2$,
which is used in Ref.~\cite{LHCb:2018vuc}.}, however, there is no sign of any enhancement in the 
data\footnote{A different SU(3) assignment, { e.g.}, $(\boldsymbol{6}^{\prime},J=1)$ would require
a strong peak $\sim 100$~MeV/$c^2$ {\em below}  $\Sigma_b(6097)$ and two hyperfine split
peaks $~200$~MeV/$c^2$ {\em above}, { i.e.}  within the range scanned by LHCb.}.
A possible explanation would be, that the decay rate of the $J=1$ multiplet to the ground state 
$\overline{\boldsymbol{3}}$ and a pseudoscalar meson
 is suppressed by $1/m_b^2$ (see Table~\ref{tab:dcouplings}). This suppression is too small
in the charm sector, this is why the excited $\Omega_c$ states have been found with small, albeit
observable widths. On the other hand $p\,$-wave decay to the excited antitriplet is not suppressed, so
$\Sigma_b(6097) \rightarrow \Lambda_b'(5912,\, 5920) +\pi^{\pm}$ should be visible. 
$\Sigma_b$ states in $\boldsymbol{6}^{\prime}(J^{P}=2^{-}) $  are above the energy range scanned by  LHCb; however, they
are expected to be rather wide $d\,$-wave resonances.

We can roughly estimate partial decay width of $\Sigma_b(6097)$, since on general grounds 
\cite{Cheng:2015naa}
the $l\,$-wave decay width of baryon $B_1$ to $B_2$  where the soliton is in  the SU(3) representations 
$\boldsymbol{R}_{1,2}$ respectively, and a pseudoscalar meson $\Phi$, is proportional to,
\begin{equation}
\Gamma_{B_1\rightarrow B_2+\Phi}\sim \frac{p_{\Phi}^{2l+1}}{F_{\Phi}^2} \frac{M_2}{M_1}
\left[
\begin{array}
[c]{cc}%
8 & \boldsymbol{R}_{2}\\
\Phi & B_{2}%
\end{array}
\right\vert \left.
\begin{array}
[c]{c}%
\boldsymbol{R}_{1}\\
B_{1}%
\end{array}
\right] ^{2} \, .
\label{eq:Gammaratio}
\end{equation}
Here $M_{1,2}$ are masses of baryons $B_{1,2}$, $F_{\Phi}$ is $\Phi$ meson decay constant,
and the square bracket corresponds to the
pertinent SU(3) isoscalar factor. From Eq.~(\ref{eq:Gammaratio}) we get
\begin{equation}
\frac{\Gamma(\Sigma_b(6097)\rightarrow\Lambda_b(5620)+\pi)}
{\Gamma(\Omega_c(3000)\rightarrow \Xi_c(2469)+\overline{K})}
\sim 2.25\, ,
\end{equation}
since both decays are in $s\,$-wave, and both $\Sigma_b$ and $\Omega_c$ are
assumed to belong to 
$(\boldsymbol{6}^{\prime},J=0)$.
Given rather small width of $\Omega_c(3000)$ 
of $\sim 4.5$~MeV, we get 
\begin{equation}
\Gamma(\Sigma_b(6097)\rightarrow\Lambda_b(5620)+\pi) \sim 10~\text{MeV}\,.
\label{eq:Sigmawidth}
\end{equation}
LHCb estimated the total width to be 30~MeV, however, theoretical total width will be larger than 10~MeV, as the new channels,
like $\Sigma_b(6097)\rightarrow\Lambda^{\prime}_b+\pi$ open.

In the charm sector the  isotriplet of excited baryons decaying into $\Lambda_c^+ +\pi$ was observed in 2005
by the Belle Collaboration~\cite{Belle:2004zjl} with mass of 2800~MeV/$c^2$ and is now included in the
PDG \cite{PDG} (see Table~\ref{tab:unknown_c}). Originally Belle has tentatively identified these resonances
as $S^P=3/2^-$ but from the $J=2$ multiplet. In our Fit 1 $\Sigma_b(1800)$ would be $S^P=1/2^-$, in Fit 2 is $S^P=3/2^-$ but from $J=1$ sextet.

Belle has scanned rather large range of  $\Lambda_c^+ +\pi$ invariant mass from 2285~MeV/$c^2$ to 3085~MeV/$c^2$,
where all five $\Sigma_c$ resonances should be seen. On the other hand the Babar Collaboration
in a similar mass region reported a $\Sigma_c$ state at 2846~MeV/$c^2$~\cite{BaBar:2008get}, 
which is $3\sigma$ from the Belle
measurement, or a new resonance. Certainly experimental situation in this region is not clear, moreover
one should remember that in Babar 
 charmed baryons are produced from $B$ decays, and therefore their yields depend on the production
 mechanism. It is therefore possible that $B$ decays to higher-spin baryons are suppressed~\cite{BaBar:2008get}.

\subsection{Where are the missing $\Xi_Q$ states?}

In 2018 the LHCb Collaboration reported new $\Xi_b^{0,-}(6227)$ states in the invariant mass distributions of $\Lambda_b \overline{K}$ and $\Xi_b\pi$
over the range 6120 -- 6520~MeV$/c^2$
of total width $\Gamma\sim 18.1\pm 5.7$~MeV~\cite{LHCb:2018vuc}. It follows from our discussion in Sec.~\ref{subs:6prim} that
$\Xi_b(6227)=\Xi_b(\boldsymbol{6}',J^P=0^-)$. In the same energy range we expect two hyperfine split $\Xi_b(6327)$ and $\Xi_b(6333)$,
which we have classified as $(\boldsymbol{6}',J^P=1^-)$. These states have been recently discovered by the LHCb Collaboration \cite{LHCb-talk}
in a three body decay to $\Lambda_b^0 K^-\pi^+$, however, LHCb suggested a different interpretation following Refs.~\cite{Chen:2018orb,Chen:2019ywy}, 
namely as a $1D$ doublet of spin
$3/2^+$ and $5/2^+$. On the other hand, according to the present interpretation, for these states two-body decays are suppressed as $1/m_b^2$ and therefore
are not seen in $\Lambda_b \overline{K}$ and $\Xi_b\pi$ mass distribution. Furthermore, as seen from Tables~\ref{tab:test0}--\ref{tab:test3}, two
hyperfine split $(\boldsymbol{6}',J^P=2^-)$ states are above the energy range scanned by LHCb.

As in the case of $\Sigma_b(6097)$ we shall try to estimate the decay width  of $\Xi_b(6227)$. From Eq.~(\ref{eq:Gammaratio}) we obtain for $s\,$-wave
decays
\begin{equation}
\frac{\Gamma(\Xi_b(6227) \rightarrow \Xi_b(5795)+\pi)}{\Gamma(\Xi_b(6227) \rightarrow \Lambda_b(5620)+K)}\sim 2.7 \, .
\end{equation}
On the other hand 
\begin{equation}
\frac{\Gamma(\Xi_b(6227) \rightarrow \Xi_b(5795)\pi)}{\Gamma(\Sigma_b(6097)\rightarrow\Lambda_b(5620)+\pi)} \sim 0.7 \, .
\end{equation}
Therefore the total width of $\Xi_b(6227)$ can be estimated as 
\begin{align}
&\Gamma(\Xi_b(6227)  \rightarrow \Xi_b(5795)+\pi) \notag \\ 
&+\Gamma(\Xi_b(6227) 
 \rightarrow \Lambda_b(5620)+K) \sim 9.6~\text{MeV}
\end{align}
where we have used (\ref{eq:Sigmawidth}). This is almost two times too small, but still within the accuracy of the present model, given 
large experimental error for the total $\Xi_b(6227)$ width of 5.7~MeV. Note that, unlike in the case of $\Sigma_b(6097)$, 
$\Xi_b(6227)$ has no open channel to $\overline{\boldsymbol{3}}'+$~pseudoscalar meson.

\subsection{Comparison with the Skyrme model}
\label{ssec:Skyrme}

The approach to heavy baryons in the soliton models dates back to the late 1980s and was initiated by a seminal paper by Callan
and Klebanov \cite{Callan:1985hy} (with subsequent improvement in \cite{Callan:1987xt}) where strange baryons are viewed as
bound-states of kaonic fields in the soliton background of the Skyrme model \cite{Skyrme:1961vq,Adkins:1983ya}. 
Subsequently, with an advent of heavy quark symmetry \cite{Isgur:1991wq}, soliton bound-states with charm or beauty
mesons have been used to describe heavy baryons \cite{Jenkins:1992zx}.

There are certain similarities and differences between our approach and the one of the Skyrme model. First of all the soliton
in our picture is a relativistic quark configuration (see Fig.~\ref{fig:levels}) with valence levels and a fully occupied Dirac sea \cite{Diakonov:1987ty}. 
The energy of the Dirac sea can be approximated using gradient expansion, yielding expressions analogous to the Skyrme
lagrangian \cite{Diakonov:1987ty}. In an artificial limit where the soliton size is increased beyond its physical value that minimizes
the aggregate energy of the valence level and the Dirac sea, the valence level sinks into the negative continuum and the model
resembles the Skyrme model. The soliton energy in this case can be expressed entirely in terms of the gradient expansion 
of the chiral field. However, in the
realistic situation the quark degrees of freedom are more appropriate for heavy baryon description. It would be interesting to see how in the
large soliton limit the hole in the Dirac sea (from the missing valence quark - see~Fig.~\ref{fig:levels}.b) combines with the heavy quark to form 
an effective heavy meson field.

On phenomenological side it is difficult to make an exact comparison of the bound-state Skyrme model results and our predictions,
because - to the best of our knowledge - in all Skyrme model calculations strangeness was also included in terms of the kaonic 
bound-states. On the  contrary, we treat the strange quark mass perturbatively and rely rather heavily on the underlying SU(3)
representation structure, which is missing in the bound-state approach.

Nevertheless, some comparison is possible if we confine ourselves only to the SU(2) substructure. The authors of Refs.~\cite{Min:1992uk,Oh:1994vd}
computed rotational and $1/m_Q$ corrections to the bound-state approach that lead to the hyperfine splittings of different spin multiplets.
Expanding their \cite{Min:1992uk} formula (28) in terms of a constant $c \sim 1/m_Q$ one recovers rotational energy analogous to
our formula (\ref{rotmass}) and, in the linear order in $c$, spin-spin interaction identical to our 
Eq.~(\ref{eq:ssinter}) (with obvious exchange of their unknown parameter $c$ by $\kappa/m_Q$).\footnote{Note that $\bm{J}\cdot\bm{J}_{Q}=(S(S+1)-J(J+1)-J_Q(J_Q+1))/2$, and $J_Q(J_Q+1)=3/4$, where $S$ and $J$ 
denote baryon spin and soliton angular momentum, respecively.}

The bound-state approach boils down to finding a solution to the corresponding Schrödinger equation for a meson-soliton bound-state 
in a potential provided by the soliton background. These modes correspond to the $\lambda$ modes discussed in the Introduction. Typically
the soliton potential is rather weak allowing only for a few such states. For example in Ref.~\cite{Cohen:2007up} only two states with
orbital momentum  $l=0$ have been found and one with $l=1$. No states with higher $l$'s have been observed. This observation,
although obviously strongly model dependent,  reinforces
our approach where $\lambda$ modes have been neglected.

\section{Conclusions}

\label{sec:conclusions}

In this paper we have tried to classify recently discovered heavy baryons in terms of the SU(3) multiplets of the light
subsystem. To this end we have used heavy quark symmetry and the chiral quark-soliton model for which one can derive
mass formulas both for the ground state baryons and for excited states. The relative heavy quark-soliton excitations,
so called $\lambda$-modes, have not been taken into account. We have presented arguments that they are parametrically suppressed in the
large $N_c$ limit, on which the $\chi$QSM is based. For the purpose of this analysis we have adopted so called { model-independent}
approach, where {\em a priori} calculable quantities are fitted from a small number of input masses. The remaining masses and splittings
can be then predicted.

Although the mass formulas (\ref{eq:6massesg}) and (\ref{eq:6massese}) are the same both for charm and beauty baryons, 
it is clear that the heavy quark symmetry should work better in
the $b$ sector. Therefore as the only input from the charm sector we used the SU(3) assignment \cite{Kim:2017jpx}
of excited $\Omega_c$ states discovered by the LHCb
Collaboration in 2017 \cite{LHCb:2017uwr}.
This assignment has been further reinforced by the study of the decay widths of some of the $\Omega_c$ states \cite{Kim:2017khv}.

Further input came  from the $b$ sector (\ref{eq:binputs}) based on the hyperfine splittings and the constraints from the $c$ sector.
Our results are  best illustrated in Table~\ref{tab:test1}. We have shown that all known $\Xi_b$ and $\Sigma_b$ states from
Table~\ref{tab:unknown_b} can be interpreted as members of different sextets of negative-parity. Unfortunately recently reported 
$\Omega_b$ states pose a problem, especially their mass differences that are much smaller than the $\Omega_c$ mass splittings
in the charm sector, see however Ref.~\cite{Karliner:2020fqe}.

In the charm sector our analysis leaves two unassigned  states: $\Xi_c(3056)$ and $\Xi_c(3079)$. It has been observed in 
Refs.~\cite{LHCb:2020tqd,Bijker:2020tns,gaps_Chen} that the following mass splittings are equal:
$
\Omega_c(3050)-\Xi_c(2923),~\Omega_c(3065)-\Xi_c(2939),~\Omega_c(3090)-\Xi_c(2965)
$, leading to the supposition that all these states belong to the SU(3) sextets. In our case
the  equality of the first splitting with the two others is rather accidental, as 
$\Omega_c(3050)$ and $\Xi_c(2923)$
in our
model belong to the exotic $\overline{\bf 15}$. Nevertheless the mass difference is numerically approximately equal to
the ones in the $J=1$ sextet (see Table~III in Ref.~\cite{Kim:2017jpx}). Moreover, splittings in excited sextets of different $J$ are 
not equal in the present approach [see Eqs.~(\ref{eq:6massesg}) and (\ref{eq:6massese})], although numerically the difference may be quite small.

With the proposed assignment we have analyzed possible two body decay patterns giving arguments why some states have not been seen
in the two-body mass distributions. We have also predicted masses of yet unmeasured members of the excited sextets. 


\section*{Acknowledgments}
This paper has been started in collaboration with Maxim V. Polyakov from the Ruhr-University in Bochum. We profited
a lot from discussions with Victor Yu. Petrov from the Institute of Nuclear Physics in Gatchina, St. Petersburg. Both of them passed away in August
and September 2021 respectively, therefore this article is a modest attempt to bring all the discussions and ideas together 
by the present author (MPr). This paper is a continuation of  earlier works done in collaboration with Hyun-Chul Kim and Ghil-Seok Yang, whom we 
are indebted for discussions and hospitality.
MPr has been supported by the Polish National Science Centre Grants 2017/27/B/ST2/01314
and 2018/31/B/ST2/01022.
MPr thanks also the Institute for Nuclear Theory at the University of Washington for its kind hospitality and stimulating research environment. This research was supported in part by the INT's U.S. Department of Energy Grant No. DE-FG02- 00ER41132.

\end{document}